\documentclass[reprint,amsmath,amssymb,aps,prresearch,longbibliography]{revtex4-2}
\usepackage{graphicx}
\usepackage{bm}
\usepackage{physics}
\usepackage{hyperref}

\begin{document}

\title{Inferring Microscopic Explanatory Structures from Observational Constraints via Large Deviations}
\author{Akihisa Ichiki}
\email{ichiki@fukuoka-u.ac.jp}
\affiliation{Department of Applied Mathematics, Faculty of Science, Fukuoka University, 8-19-1, Nanakuma, Jonan-ku, Fukuoka City, 814-0180, Japan}

\date{\today}

\begin{abstract}
We study how macroscopic observational constraints restrict admissible microscopic explanatory structures when no intrinsic order or dynamics is assumed a priori. 
Starting from an unordered collection of measurement outcomes, we formulate inference as a constrained large deviation problem, selecting probability assignments that minimize relative entropy with respect to a reference measure determined solely by the measurement setup. 
We show that among all microscopic structures compatible with a given macroscopic constraint, those rendering the observation statistically most typical are selected. 
As an explicit illustration, we demonstrate how ordered microscopic structures can emerge purely from inference under constraint, even when the reference measure is fully permutation symmetric. 
Order is thus not assumed but inferred, serving here only as an illustrative example of a broader class of relational explanatory hypotheses constrained by observation. 
\end{abstract}

\maketitle

\section{Introduction}

Observation rarely presents itself in an ordered form. 
Measurements yield collections of outcomes, often subject to redundancy, coarse-graining, and symmetries inherent to the measurement setup. 
From such data, structure is not given but inferred. 
The problem addressed here is therefore not to identify a unique microscopic model, but to characterize the space of microscopic explanatory hypotheses compatible with given macroscopic observations. 
Accordingly, in this work, order serves only as an illustrative example of a relational explanatory structure; the central issue is how observational constraints delimit admissible microscopic hypotheses. 
We illustrate this perspective by considering microscopic order inferred from macroscopic observation. 
Then a fundamental question arises: 
\emph{Can relational explanatory structures be inferred from unordered observations, without assuming any intrinsic organization of the underlying events?} 

In many theoretical frameworks, order is introduced implicitly. 
Sequential indices, temporal parameters, or causal relations are assumed before statistical inference begins. 
While such assumptions are often operationally justified, they obscure the extent to which order itself may be an emergent construct. 

In this work, we deliberately adopt a more primitive standpoint~\cite{jaynes1957information}. 
We assume that the observer is given only an unordered collection of measurement outcomes. 
These outcomes arise from underlying events through measurements performed under local gauge realizations, which limit what can be observed and induce nontrivial statistical structure~\cite{caticha2012entropic}. 
No structure---order or otherwise---is assumed at the outset. 

Macroscopic observations enter our framework as constraints: 
they specify what must be explained, but not how individual observations are related. 
To select among competing explanatory structures, we employ a large deviation principle~\cite{TOUCHETTE20091}, minimizing relative entropy with respect to a reference measure determined solely by the measurement setup. 

Within this approach, order appears not as a primitive notion, but as a hypothetical explanatory structure. 
Among all candidate structures compatible with the macroscopic constraint, the one that renders the observation statistically most typical is selected. 
Crucially, this selection can break permutation symmetry even when the reference measure itself is fully symmetric. 

The aim of this paper is to demonstrate this mechanism explicitly. 
We show that ordered structures can emerge purely from inference under constraint, without invoking dynamics, causality, or sequential data
acquisition. 
It is worth noting that order is treated here as one illustrative relational structure. 
To make the argument transparent, we construct a minimal model based on binary events. 

The paper is organized as follows. 
Section~\ref{sec:2} formalizes the notion of observation in the absence of intrinsic relational structure and introduces the measurement-induced reference measure. 
Section~\ref{sec:3} introduces hypothetical relational structures and the class of models used to explain macroscopic observations. 
Section~\ref{sec:4} summarizes the relevant aspects of large deviation theory. 
Section~\ref{sec:5} presents the core argument showing how relational structure such as order is selected by entropy minimization. 
Section~\ref{sec:6} illustrates the mechanism using a minimal binary model. 
Section~\ref{sec:7} concludes the paper by summarizing how observational constraints restrict admissible microscopic explanations and by discussing the scope and limitations of this inference-based perspective. 

\section{Observation and Reference Measure from Measurement Symmetry}\label{sec:2}

We begin by specifying what is meant by an observation, without assuming any notion of structure, such as temporal or causal order. 

Let $e$ denote an event occurring in the system. 
An observer does not access the event directly but instead records a measurement outcome or label $x \in \mathcal{X}$, which depends both on the event and on the choice of measurement reference. 
We represent this as 
\begin{equation}
x = x(g, e),
\end{equation}
where $g \in G$ denotes a transformation associated with the measurement setup, such as the choice of origin, calibration, or reference frame. 
Here $G$ denotes a group of transformations, which exchanges measurement references. 
Different choices of $g$ correspond to different but equally admissible measurement conventions. 

The same event $e$ may therefore produce different labels depending on $g$: 
\begin{equation}
g_1 x := x(g_1 g_0, e) \neq x(g_0, e) = x.
\end{equation} 
Conversely, for a given label $x$, there generally exist multiple pairs $(g, e)$ that realize it. 
In this sense, $x$ should be understood as a \emph{microscopic observational state}, not as an intrinsic property of the event itself. 

\subsection{Measurement-induced reference measure}

The measurement setup determines which labels are accessible and how frequently they appear under changes of the measurement reference. 
We assume that the observer has no prior information about the events themselves and only knows the measurement protocol. 

This induces a reference probability measure $\pi$ on $\mathcal{X}$, defined by the structure of the $G$-action~\cite{wong2024gauge,
rovelli1996relational, sanchez2008foundations, henneaux1992quantization, BELOT2003189}: 
\begin{equation}
\pi(x) \propto \dfrac{1}{|\mathcal{O}_x|},
\end{equation}
where $\mathcal{O}_x = \{g \in G | x = x(g, e)\mbox{ for some }e\}$ denotes the orbit associated with the label $x$, and $|\mathcal{O}_x|$ is the number of its elements. 
Here, possible degeneracies arising from distinct events producing the same label are absorbed into an uninformative prior over events, leaving a reference measure determined solely by the symmetry of the measurement protocol. 
In other words, $\pi$ is fixed once the measurement scheme is fixed, prior to any observation of data. 

We emphasize here that $\pi$ is not an empirical distribution inferred from data. 
It represents the observer’s reference measure associated with the microscopic labeling procedure itself. 
This reference measure reflects the symmetry of the measurement protocol rather than any assumption about the system itself. 

\subsection{Observed data}

An experiment produces a finite collection of microscopic labels 
\begin{equation}
\omega = \{ x_1, \cdots, x_N\},
\end{equation}
corresponding to $N$ measurement outcomes. 

Crucially, we do not assume that these labels are observed in any meaningful order. 
The object $\omega$ is therefore a multiset, or equivalently an element of $\Omega_N := \mathcal{X}^N/S_N$, where permutations of indices are identified. 
Here $S_N$ denotes a permutation group on a set consisting of $N$ elements. 

The absence of order is not a modeling choice but a statement about the observational situation:
no temporal, causal, or sequential information is assumed to be directly accessible at this stage. 

Given the reference measure $\pi$, the corresponding reference probability on $\Omega_N$ is 
\begin{equation}
Q(\omega) = \displaystyle\prod_{i=1}^N \pi(x_i),
\end{equation}
which represents the observer’s state of minimal information about the observed microscopic data~\cite{cohn2013measure, cox1979theoretical, jordan2024quantum}. 

\section{Symmetry and Hypothetical Explanatory Structures}\label{sec:3}

The collection of microscopic outcomes $\omega$ by itself carries no notion of sequence. 
Nevertheless, macroscopic observations often suggest regularities that are naturally expressed in terms of ordered structures~\cite{caticha2012entropic}. 

To address this, we introduce order not as an assumption, but as a \emph{hypothetical explanatory structure}. 
We begin by clarifying the symmetry properties of the microscopic configuration $\omega$ and the role of the observational label $\sigma \in S_N$. 

\subsection{Permutation invariance of $\omega$}

Let $\sigma \in S_N$ be a permutation of the $N$ observed labels. 
Given $\sigma$, the unordered set $\omega$ may be represented as an ordered sequence 
\begin{equation}
(\omega, \sigma) := (x_{\sigma(1)}, \cdots, x_{\sigma(N)}).
\end{equation}
Notice here that $\sigma$ is not observable. 
It does not correspond to a measured time or causal relation. 
It represents a hypothetical ordering introduced for the purpose of explanation. 
In other words, the multiset $\omega = \{ x_1, \cdots, x_N\}$ has an invariance property, 
\begin{equation}
\sigma \omega := \{ x_{\sigma(1)}, \cdots, x_{\sigma(N)} \} = \omega, \qquad \forall \sigma \in S_N.
\end{equation}
This invariance is not an assumption but a defining property of the configuration space. 
The permutation group $S_N$ therefore does not act nontrivially on $\omega$. 

\subsection{Models for macroscopic observations}

Suppose a macroscopic observable $m$ is experimentally accessible. 
At the level of events, it is assumed to depend only on the underlying microscopic events and not on the measurement reference: 
\begin{equation}
m = \tilde{M}(e_1, \cdots, e_N).
\end{equation}
However, the observer has access only to the microscopic labels $\omega$. 
Accordingly, the macroscopic observation must be explained by a model of the form 
\begin{equation}
m = M(\omega, \sigma)/N,
\end{equation}
where $M$ is a function of the unordered microscopic data and a hypothetical ordering, extensive to the size of $\omega$. 
From this perspective, $M$ should be regarded as an \emph{explanatory model} rather than a directly observable quantity. 

\subsection{Observables and constraint independence}

We introduce an observable $M(\omega,\sigma)$ that depends explicitly on an ordering $\sigma \in S_N$. 
While $M$ itself may vary with $\sigma$, the physical constraint imposed through observation is formulated at the level of expectation values. 

Specifically, we require that 
\begin{equation}
\langle M \rangle_P / N = m
\label{eq:constraint}
\end{equation}
be independent of the particular choice of $\sigma$, where the expectation value is defined as 
\begin{equation}
\langle M\rangle_P
:= \sum_\omega P(\omega) M(\omega,\sigma).
\end{equation}
Here $P$ denotes a probability measure over configurations $\omega$. 
This requirement states that, under a relabeling $\tau\in S_N$, 
\begin{eqnarray}
P(\omega)&\mapsto& P^\tau(\omega)=P(\tau\omega)=P(\omega),\nonumber\\
M(\omega,\sigma)&\mapsto& M^\tau(\omega,\sigma)=M(\omega,\tau\sigma),
\end{eqnarray}
the expectation value $\langle M\rangle_P$ is invariant, 
\begin{equation}
\langle M^\tau\rangle_{P^\tau}=\langle M\rangle_P.
\end{equation}
This requirement expresses the principle that macroscopic observational constraints must not depend on the internal labeling used to probe the system. 

In addition, although neither the probability measure $P(\omega)$ nor the explanatory model $M(\omega,\sigma)$ is required to be invariant under the gauge action of measurement references $G$, the constraint imposed by macroscopic observation must be $G$-invariant. 
Specifically, under a change of measurement references $g \in G$, 
\begin{eqnarray}
P(\omega)&\mapsto& P^g(\omega)=P(g\omega),\nonumber\\
M(\omega,\sigma)&\mapsto& M^g(\omega,\sigma)=M(g\omega, \sigma),
\end{eqnarray}
the expectation value $\langle M\rangle_P$ is invariant, 
\begin{equation}
\langle M^g\rangle_{P^g}=\langle M\rangle_P,
\end{equation}
where $g\omega := \{ gx_1, \cdots, gx_N\}$. 
This expresses the fact that macroscopic constraints are insensitive to arbitrary measurement references, which leads relabelings of microscopic events. 

At this stage, no assumption is made as to whether such a constraint can be satisfied nontrivially. 
We show in later sections that enforcing such a gauge-invariant constraint leads to a nontrivial selection among hypothetical structures. 

\section{Large Deviation Theory as an Inference Tool}\label{sec:4}

We now introduce the mathematical principle~\cite{jaynes1957information, TOUCHETTE20091, dembo1998large, ellis2012entropy} used to select plausible explanations. 

Given the reference probability $Q(\omega)$, we seek a probability measure $P(\omega)$ that incorporates macroscopic information while remaining as close as possible to $Q$. 

\subsection{Relative entropy and constrained inference}

The relative entropy (Kullback--Leibler divergence) is defined as 
\begin{eqnarray}
D(P\|Q) = \displaystyle\sum_{\omega}P(\omega)\ln\dfrac{P(\omega)}{Q(\omega)}\,.
\end{eqnarray}
For a fixed value of $\sigma$, we define a probability measure $P^*_\sigma(\omega)$ as the solution of the variational problem 
\begin{equation}
P^*_\sigma
=
\arg\min_P
\left\{
D(P\|Q)
\;\middle|\;
\langle M(\omega,\sigma)\rangle_P/N = m
\right\},
\label{eq:varP}
\end{equation}
where $Q(\omega)$ is the reference measure determined by the observational setup~\cite{PhysRevE.84.061113, YukiSUGHIYAMA2013IIS190114}. 
Following the Cram\'er-type large deviation principle~\cite{cramer1938}, $P^*_\sigma$ is characterized as the most probable distributions compatible with macroscopic constraints with fixed $\sigma$. 

The solution of \eqref{eq:varP} takes the exponential reweighting form 
\begin{equation}
P^*_\sigma(\omega)
=
\frac{1}{Z(\sigma)}
Q(\omega)\exp\!\left[\lambda\, M(\omega,\sigma)\right],
\label{eq:Psigma}
\end{equation}
where $\lambda$ is a Lagrange multiplier determined by the constraint and $Z(\sigma)$ is the normalization factor. 
The Lagrange multiplier $\lambda$ is non-negative, ensuring that larger values of $M(\omega,\sigma)$ correspond to higher probability under
$P^*_\sigma$. 

\subsection{Role of large deviations in the present framework}

In the present context, a large deviation theory is not used to model dynamics or time evolution. 
By conrast, it provides a principled way to infer which microscopic descriptions are most consistent with macroscopic observations. 

It is worth noting that the constraint involves the hypothetical orderings $\sigma$. 
As a result, the inference problem implicitly compares different hypotheses and selects those that best reconcile the macroscopic constraint with the reference measure. 
This mechanism will be shown to lead to the \emph{selection of an effective explanatory structure} in the next section. 

\section{Hypothesis Selection by Constrained Large Deviations}\label{sec:5}

In this section, we show how an explanatory structure is selected from observational data without assuming any prior structure. 
The variational principle introduced in the previous section with the requirements stated in Sec.~\ref{sec:3} leads naturally to the selection of explanatory structure in the large-$N$ limit. 

\subsection{Structure selection by realized observation}

We consider a single observational outcome $\omega_{\mathrm{obs}}$ given as an unordered multiset of microscopic labels. 
No intrinsic structure is assumed at this stage. 

To explain the observed macroscopic quantity $m$, we introduce a family of hypothetical explanatory structures represented by a function $M(\omega,\sigma)$. 
While the variational principle determines $P^*_\sigma$ for each fixed $\sigma$ as discussed in the previous section, the actual observational data provide only a single realization $\omega_{\mathrm{obs}}$. 
The explanatory structure must therefore be selected according to how well the induced measure $P^*_\sigma$ accounts for this realization. 
In this setup, $\sigma$ labels hypothetical relational structures (e.g. orderings). 

To compare the inferred measures $P^*_\sigma$, the relative entropy is again emploied. 
A probability for observing an empirical measure $P$ of samples from a population with a probability measure $P_0$ is given by Sanov's theorem: 
\begin{equation}
{\rm Prob}(P) \asymp \exp[-ND(P\|P_0)].
\end{equation}

Given an observed realization $\omega_{\rm obs}$, we consider the empirical measure $P_{\rm emp}(\omega) = \delta(\omega-\omega_{\rm obs})$. 
Treating $P^*_\sigma$ as a reference measure, Sanov’s theorem~\cite{sanov1958probability} implies that the probability of observing $P_{\rm emp}$ under the hypothesis $\sigma$ scales as 
\begin{eqnarray}
{\rm Prob}(P \approx P_{\rm emp})&\asymp& \exp[-N I(\sigma)],\nonumber\\
I(\sigma)&=&D(P_{\rm emp}\| P^*_\sigma).\label{eq:Sanov}
\end{eqnarray}
The rate function $I(\sigma)$ can be explicitly written as 
\begin{equation}
I(\sigma) = \displaystyle\sum_\omega P_{\rm emp}(\omega)\left[ \ln P_{\rm emp}(\omega) - \ln P^*_\sigma(\omega)\right],
\end{equation}
where the first term in the right-hand side is constant with respect to $\sigma$, and the second term equals $-\ln P^*_\sigma(\omega_{\rm obs})$. 
Thus we select $\sigma$ by maximizing the probability assigned to the observed outcome: 
\begin{eqnarray}
\sigma^* &=&
\arg\max_\sigma P^*_\sigma(\omega_{\mathrm{obs}})\nonumber\\
&=&
\arg\max_\sigma\left[\lambda\, M(\omega_{\mathrm{obs}},\sigma) - \log Z(\sigma)\right].
\label{eq:effectivesigma}
\end{eqnarray}

Equation (\ref{eq:effectivesigma}) shows that the explanatory structure is selected by a competition between the fit to the realized observation, measured by $M(\omega_{\mathrm{obs}},\sigma)$, and a normalization term reflecting the typicality of outcomes under the induced measure. 

As a result of large deviation principle \eqref{eq:Sanov}, in large $N$ limit, enforcing a $\sigma$-independent expectation value selects a particular hypothesis $\sigma^*$ that dominates the measure. 
The symmetry of the constraint is thus preserved, while the solution that realizes it is not. 

If, for a given observational realization, the maximization (\ref{eq:effectivesigma}) does not select a symmetry-breaking $\sigma$, then no specific hypothesis is selected even at the explanatory level. 
From this perspective, the framework allows for the possibility that hypothetical structures are neither present in the data nor required for their
explanation. 

This mechanism constitutes a selection principle driven purely by observational consistency, rather than by explicit symmetry breaking at the level of assumptions. 

It is important to note that $\min_\sigma I(\sigma)$ need not vanish. 
In such cases, the observed realization is exponentially unlikely under all hypotheses $\sigma$, indicating either a misspecification of the explanatory observable $M$ or an intrinsic inadequacy of the pair $(\omega, \sigma)$ to account for the macroscopic observation $m$. 

\subsection{Relation to regression and maximum likelihood}

Formally, the selection rule \eqref{eq:effectivesigma} is a form of maximum likelihood estimation. 
The explanatory structure $\sigma$ is chosen so as to maximize the probability assigned to the observed outcome under the induced measure
$P^*_\sigma$. 

The distinction from standard regression lies not in the inference principle itself, but in what is being inferred. 
In conventional regression or likelihood-based inference, one typically fits numerical parameters within a fixed parametric family~\cite{cox2001algebra, amari2016information}. 
Here, by contrast, the parameter $\sigma$ labels elements of a group or relational structure, such as permutations or symmetry transformations. 

From this perspective, the present framework can be viewed as a group-based or symmetry-based extension of maximum likelihood inference, in which explanatory structures rather than numerical parameters are selected. 
Crucially, the likelihood function itself is not postulated but induced by a variational principle on probability measures. 

\section{A Minimal Binary Model: Constraint and Hypothesis Selection}\label{sec:6}

We illustrate our viewpoint---that macroscopic observations constrain and sometimes reject microscopic models---by a minimal combinatorial example. 

Let $\omega = \{x_1,\ldots,x_N\}$ be a multiset with elements $x_i \in \{0,1\}$. 
The ordering of elements is assumed to be unobservable, and different orderings are represented by the action of the permutation group $S_N$.

We assume that the observational setup is symmetric under relabeling $0 \leftrightarrow 1$. 
This defines a local gauge symmetry $G = S_2$ acting on each label. 
Accordingly, the reference measure on single-event labels is taken as $\pi(x)=1/2$ with $x\in\{0,1\}$. 
The corresponding reference measure on $\omega$ is the product measure $Q(\omega)=2^{-N}$. 

For a permutation $\sigma \in S_N$, we define an extensive quantity 
\begin{equation}
M(\omega,\sigma)
= \sum_{i=1}^{N-1} \mathbf{1}_{x_{\sigma(i)} = x_{\sigma(i+1)}},
\end{equation}
which counts the number of adjacent identical events in the ordered sequence specified by $\sigma$. 
While $M$ depends on the microscopic ordering $\sigma$, the observer has access only to the macroscopic quantity $m = \langle M \rangle_P / N$. 
In this sense, $M$ plays the role of a microscopic model, whereas $m$ is a macroscopic observable. 

We first consider the case where $\omega_{\rm{obs}}$ contains $0$ and $1$ in equal proportions. 
In this situation, depending on the ordering $\sigma$, the value of $M(\omega_{\rm{obs}},\sigma)$ can range from $0$ to $N-1$, and thus the normalized quantity $m$ may take any value in $[0,1)$. 
However, once a specific value of $m$ is observed, a large deviation principle selects, with overwhelming probability, a restricted set of permutations that are compatible with the observed macroscopic constraint. 
These selected permutations may be regarded as microscopic hypotheses explaining the macroscopic observation. 
Such a restricted set of permutations determines a specific ordering or relational structure in $\omega_{\rm{obs}}$ at an explanatory level. 

Next, consider the extreme case $\omega_{\rm{obs}}=(0,\ldots,0)$.
Here, for any $\sigma\in S_N$, we have $M(\omega_{\rm{obs}},\sigma)=N-1$, and therefore $m=1$ in large $N$-limit is the only admissible macroscopic value.
If an observation yields $m\neq 1$, the microscopic model itself is incompatible with the data and must be rejected.
Conversely, when $m=1$ is observed, all permutations are exactly equivalent: the macroscopic observation leaves the microscopic ordering completely undetermined, reflecting an unbroken symmetry.

Finally, suppose that the fraction of $0$ in $\omega_{\rm{obs}}$ is $1/3$. In this case, the adjacency structure implies a nontrivial lower bound, $m \in [1/3,1)$. 
Therefore, an observed value $m<1/3$ cannot be explained by any permutation $\sigma$ and leads to the rejection of the microscopic model $M$ itself. 
This demonstrates that macroscopic observations may rule out not only individual microscopic hypotheses, but entire classes of microscopic models. 

This simple example captures the central message of the present work.
Macroscopic observations do not reveal a unique ``true'' microscopic state; they select, constrain, or even eliminate microscopic hypotheses through large deviation mechanisms.
Symmetry breaking, degeneracy, and model rejection all emerge here as consequences of the observation structure rather than properties imposed at the microscopic level.

\section{Conclusion}\label{sec:7}

In this work, we investigated how macroscopic observational constraints restrict admissible microscopic explanatory structures when no intrinsic order, dynamics, or generative mechanism is assumed a priori. 
Starting from an unordered collection of measurement outcomes shaped by gauge redundancy, we introduced a measurement-induced reference measure and employed a constrained large deviation principle to infer probability assignments consistent with a given observation. 
Within this framework, relational structures such as order arise not as primitive assumptions, but as explanatory hypotheses selected by observational consistency. 

A central implication of this analysis is that symmetry breaking occurs here at the level of typical explanations rather than at the level of assumptions or constraints. 
The macroscopic constraint itself is required to remain invariant under permutations and gauge transformations, while the microscopic structure selected to account for a realized observation need not be. 
Conversely, when no hypothesis dominates in the large deviation sense, the framework correctly predicts the absence of any preferred microscopic structure. 
In this way, the approach distinguishes between symmetry preserved at the observational level and symmetry broken only in explanatory realizations. 

The minimal binary model illustrates these possibilities explicitly. 
Depending on the observed macroscopic value, microscopic hypotheses may be selected, remain degenerate, or be ruled out altogether. 
Macroscopic observations therefore do not generally identify a unique microscopic description; they delimit a space of admissible explanations. 
Model rejection, degeneracy, and effective structure selection all emerge as consequences of observational constraints rather than properties imposed at the microscopic level. 

The framework developed here is intentionally static and does not address how inferred structures relate to sequential data acquisition or dynamical evolution. 
Extending the analysis to incorporate time-resolved observations, feedback, or interaction with an environment would require additional assumptions beyond those considered in the present work. 
Nevertheless, the present results clarify the extent to which microscopic explanatory structure can be inferred from macroscopic observation alone, and where genuine underdetermination necessarily remains. 

\bibliographystyle{unsrt}
\bibliography{hyp_ref}

\end{document}